\documentclass[12pt]{article}
\usepackage{psfig,rotating}

\begin{document}

\baselineskip=25pt
\textwidth 15.0truecm
\textheight 21.0truecm
\topmargin 0.2in
\headsep 1.2cm

\title{Physical Investigation of the Potentially Hazardous Asteroid (144898) 2004 VD17 {\bf \footnote{Based on obsevations carried out at the European Southern Observatory (ESO), Chile, ESO run ID 276.C-5057}}} 


\author{F. De Luise$^{1,2}$, 
D. Perna$^{1,2}$, 
E. Dotto$^{1}$,
S. Fornasier$^{3}$,\\
I. N. Belskaya$^{4}$,
A. Boattini$^{1,2}$,
G.B. Valsecchi$^{5}$,
A. Milani$^{6}$,\\ 
A. Rossi$^{7}$,
M. Lazzarin$^{8}$, 
P. Paolicchi$^{9}$,
and M. Fulchignoni$^{3}$
} 

\maketitle

{\small
\noindent
$^1$ INAF-Osservatorio Astronomico di Roma, Italy;\\ 
$^2$ University of Rome ``Tor Vergata'', Italy; \\
$^3$ University of Paris VII {\it Denis Diderot}/ LESIA-Obs. of Paris, France; \\
$^4$ Institute of Astronomy of Kharkiv National University, Ukraine \\
$^5$ INAF-IASF, Italy; \\
$^6$ Dipartimento di Matematica - University of Pisa, Italy; \\
$^7$ ISTI-CNR, Pisa, Italy; \\
$^8$ University of Padova, Italy;\\
$^9$ Dipartimento di Fisica - University of Pisa, Italy;\\
}

\noindent
Submitted to Icarus: April 2007\\
e-mail: deluise@oa-roma.inaf.it\\
phone: +39-06-9428430
fax: +39-06-9447243; 
\\

\noindent
Manuscript pages: 32  \\
Figures: 7; Tables: 5 \\

\vspace{3cm}

{\bf Running head}: Nature of (144898) 2004 VD17 

\vspace{3cm}

\noindent
{\it Send correspondence to:}\\
De Luise Fiore\\
INAF-Osservatorio Astronomico di Roma\\   
Via Frascati, 33\\
I-00040 Monteporzio Catone (Roma)\\
Italy\\
e-mail: deluise@oa-roma.inaf.it\\
phone: +39-06-94286430\\
fax: +39-06-9447243\\

\newpage
\vspace{2.5cm}

\begin{abstract}
\noindent

In this paper we present the observational campaign carried out at ESO
NTT and VLT in April and May 2006 to investigate the nature and the
structure of the Near Earth Object (144898) 2004 VD17. \\ In spite of
a great quantity of dynamical information, according to which it will
have a close approach with {\bf the} Earth in the next century, the physical
properties of this asteroid are largely unknown.  We performed visible
and near--infrared photometry and spectroscopy, as well as
polarimetric observations.  Polarimetric and spectroscopic data
allowed us to classify 2004 VD17 as an E-type asteroid. {\bf A good agreement was also found with the spectrum of the aubrite meteorite Mayo Belwa.}  On the basis
of the polarimetric albedo {\bf($p_v=0.45$)} and of photometric data, we estimated a 
diameter of about 320 m and a rotational period of about 2 hours.\\
The analysis of the results obtained by our complete survey have shown
that (144898) 2004 VD17 is a peculiar NEO, since it is close to
the breakup limits for fast rotator asteroids, as defined by Pravec and
Harris (2000). These results suggest that a more robust structure must
be expected, as a fractured monolith or a rubble pile in a ``strength
regime'' (Holsapple 2002).

\end{abstract}

\newpage


\section{Introduction}

[TABLE 1]

The asteroid (144898) 2004 VD17 is a Near Earth Object (NEO) belonging
to the Apollo group, discovered by the LINEAR asteroid survey on the
7th of November, 2004. Its orbital parameters are summarized in
Table~\ref{tab:orbpar}. 
It will encounter the Earth several times during the next 100 years: in
2032, 2041, 2067 and 2102. Between February and May 2006, the available
astrometry {\bf did not exclude} the possibility of an impact with the
Earth at the last encounter of the sequence, the one taking place in early
May 2102{\bf. This} impact was rated 2 on the Torino Scale (TS, Binzel 2000) and in
excess of $-0.30$ on the Palermo Scale (PS, Chesley et al. 2002).  Further
observations have reduced these values to TS=0 and PS$<-3$.

Although we have {\bf a lot of} dynamical information, up to now the physical
properties of (144898) 2004 VD17 are very poorly known. These
characteristics are very important to constrain its mineralogical
composition, albedo and density, allowing us to {\bf deduce} the
internal structure.

In order to investigate the nature of this NEO and to assess its
impact hazard, in April and May 2006 we carried out photometric,
spectroscopic and polarimetric observations at the European Southern
Observatory (ESO, Chile).


\section{The Observational Campaign and Data Reduction}

[TABLE 2]

The observational campaign was performed in the framework of a
Director's Discretional Time Program (Run ID 276.C-5057). The aim of
this survey was to carry out visible and near-infrared photometry and
spectroscopy together with polarimetric measurements of (144898) 2004
VD17.

In particular, our aim was: 
\begin{itemize}
\item [\it i)] 
to determine the albedo by analysing the polarimetric
data in order to constrain size and taxonomic classification. Due to
the faintness of our target at the time of observations ($m_v \simeq
19.2$ mag), polarimetry was the only way to estimate its albedo and
thus its real size;
\item[\it ii)] 
to investigate the surface composition, looking for
spectral evidence of minerals and mixtures present on its surface, to
perform the taxonomic classification and to investigate the possible
link with the known meteorite classes;
\item[\it iii)] 
to investigate the rotational status, by analysing the
periodicity and the shape of its lightcurve;
\item[\it iv)] 
to reveal a possible satellite or companion and to
check {\bf for} the presence of cometary activity;
\item[\it v)] 
to investigate the internal structure (bulk density and
internal tensile strength), crosscorrelating all the obtained results.
\end{itemize}
In Table~\ref{tab:surv} an overview of the observations is
presented. For each run, telescopes, instruments and relative adopted
techniques are presented, together with the observational
circumstances.


\subsection{Polarimetry}

[TABLE 3]

Polarimetric observations of (144898) 2004VD17 were carried out at ESO-Paranal in service mode with the telescope VLT--UT2 \textit{Kueyen}, equipped with the FORS1 instrument (see http://www.eso.org/instruments/fors1) in polarimetric mode, from April 16 to May 30, 2006, for a total of 4 observing runs
(Table~\ref{tab:surv}). 

Linear polarimetry was obtained in the Bessel V filter at 4 angles of
the $\lambda/2$ retarder plate (0$^{o}$, 22.5$^{o}$, 45$^{o}$ and
67.5$^{o}$ with respect to celestial coordinate system), covering the
phase angle range from 26$^{o}$ to 77$^{o}$. For each retarder plate
position, 3 images were acquired with exposure time of 300 s each, for
a total exposure time on the target of 1 hour during each run. \\ The
observing procedure included the acquisition of several flat field
images, taken on twilight time without polarimetric optics in the
light path, and of at least one unpolarized standard star (taken from
the ESO list at
http://www.eso.org/instruments/fors/tools/FORS\_Std/FORS1\_Std.html)
to calibrate the instrumental polarization. The zero point of the
position angle was taken from the FORS1 user manual
(http://www.eso.org/instruments/fors/doc).

The data reduction and polarimetric parameters evaluation have been
performed following the procedure described in Fornasier et
al. (2006a). The observing conditions and the final results on the
degree of polarization $P$ and on the position angle $\theta$ of the
polarization plane, together with the polarimetric quantities $P_r$
($P_r = P * cos(2\theta_r) $) and $\theta_r$ ($\theta_r = \theta -
(\phi \pm 90^{o}) $, where $\phi$ is the position angle of the
scattering plane), are reported in Table~\ref{tab:polar}.

\subsection{Visible photometry and spectroscopy}

[TABLE 4, 5] [FIGURE 1, 2]

Visible photometric and spectroscopic data were acquired in service
mode during three different nights (see Table~\ref{tab:surv}) at
ESO-La Silla, using the 3.58 m NTT (New Technology Telescope). The
telescope was equipped with the instrument EMMI (ESO Multi-Mode
Instrument) in RILD mode (Red Imaging and Low Dispersion Spectroscopy)
(http://www.ls.eso.org/lasilla/sciops/ntt/emmi/index.html). This
{\bf instrument was equipped with a detector mosaic of two thin, back-illuminated, AR coated MIT/LL
CCDs}, with an image size of 2048 x 4096 pixels each one, a field of
view, in total, of 9.1 x 9.9 arcmin and a resolution of 0.1665
arcsec/pixel.

\begin{figure}[h]
\centerline{\includegraphics{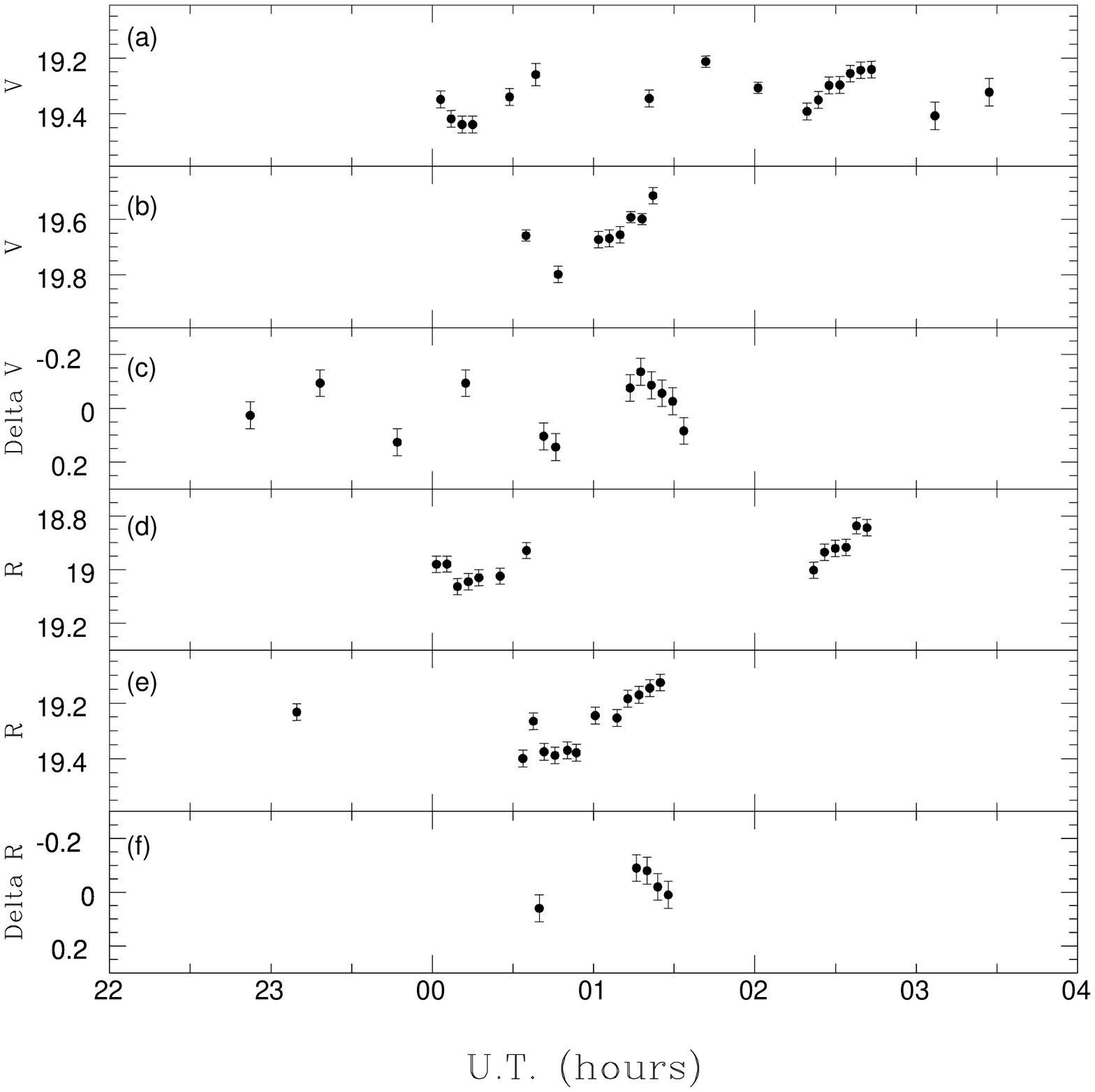}}
\caption{De Luise et al. -- Physical Investigation of PHA 2004VD17} 
\label{fig:lc_3n}
\end{figure}

For photometry the $B, V, R,$ and $I$ {\bf Johnson} filters in
$2\times2$ binning mode {\bf were used}.  {\bf The exposure times 
were} 180 and 90 sec in B
and V filters, respectively, and of 60 sec in R and I filters. These
measurements span a time interval of about four hours for the night of
1st May and about two hours in the nights of 21st and 22nd May. Biases
and sky flat-fields were also acquired at the beginning and at the end
of each night.\\ The photometric data {\bf were} reduced with the software
packages IRAF, MIDAS and IDL using the standard procedure (see, e.g.,
Dotto et al. 2006). The raw data were corrected subtracting the
bias contribution and {\bf dividing by} a median flat-field.  The
instrumental magnitude was then computed using aperture photometry,
with an integration radius of about three times the average seeing.
The sky contribution was estimated using an annulus 5-10 pixels wide
around the asteroid and then removed.  The final magnitude calibration
{\bf was} performed using several standard stars (Landolt 1992)
observed during each night, with the exception of the night of 22nd
May, that was clear but not photometric. {\bf For images acquired during this last night, a differential magnitude analysis was performed}. 
\\ The {\bf visual inspection} of the images
does not show the presence of companions of 2004 VD17, nor cometary
activity.  The obtained single night lightcurves corrected for light-time are shown in Fig.~\ref{fig:lc_3n}.  The $B$-$V$, $V$-$R$ and $V$-$I$ colors are
reported in Table~\ref{tab:cols}.

\begin{figure}[h]
\centerline{\includegraphics{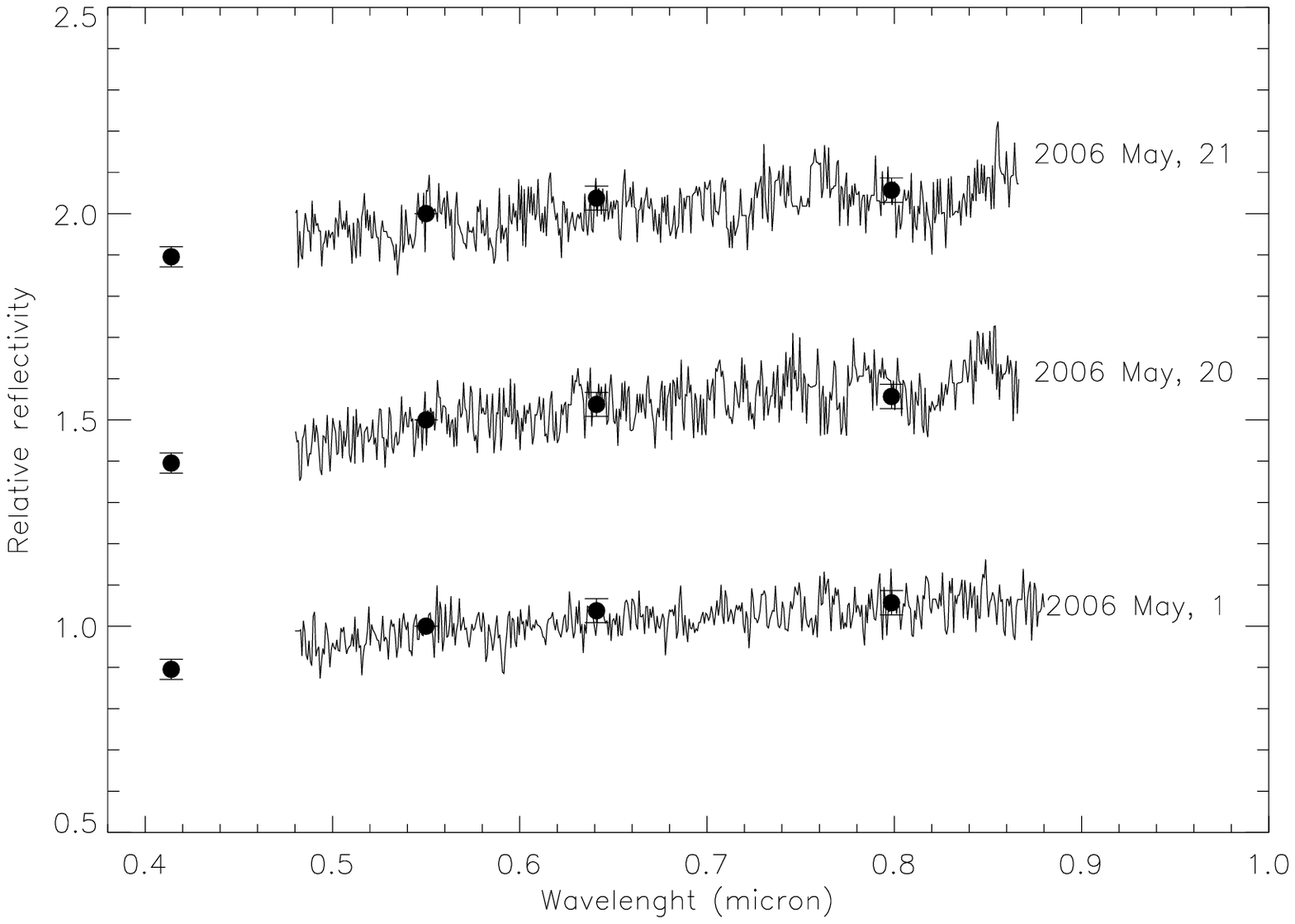}}
\caption{De Luise et al. -- Physical Investigation of PHA 2004VD17}
\label{fig:3vis}
\end{figure}

For spectroscopy {\bf was used} the grism \#1 (150 gr/mm), covering the
wavelenght range $4100\div9400$ \AA , with a dispersion of 3.1 \AA/pix
(200 \AA/mm) at the first order. The slit of 5 arcsec was oriented
along the direction of the asteroid's motion.
To reduce the possibility that the target went out of the slit during the exposure, each spectrum was divided into 2--3 parts, with a total exposure time
of 20--30 minutes (see Table~\ref{tab:texp_spc}). {\bf The asteroid position in the slit was checked before each spectrum.}\\ The obtained
spectra were calibrated in wavelenght, using a helium-argon lamp as
reference. Then the object's spectra {\bf were} divided by the spectrum
of a solar analog star observed just before and after the target, at a
similar airmass (Table~\ref{tab:texp_spc}). All the obtained visible spectra, shown in Fig.~\ref{fig:3vis}, are flat and featureless. They exhibit very
similar behaviors and have the same spectral slope of $(0.03 \pm
0.01)$ /$10^{3}$\AA$^{-1}$, computed between 0.55 and 0.75$\mu$m.


\subsection{Near-Infrared photometry and spectroscopy}

[FIGURE 3]

The near-infrared observations were carried out at ESO-Paranal in
service mode, during two nights (24th and 25th of May), using the 8.2
m VLT--UT1 \textit{Antu}, equipped with the infrared-cooled grating
spectrometer ISAAC (Infrared Spectrometer And Array Camera)
(http://www.eso.org/instruments/isaac) and with a Rockwell Hawaii
1024$\times$1024 pixel Hg:Cd:Te array.

Photometric $J$, $H$, and $Ks$ measurements (centered at 1.25, 1.65, and
2.16~$\mu$m) have been obtained before each spectrum, with an exposure
time of 120~s, 180~s and 240~s respectively.  The calibration was
performed by the observation of several faint infrared standard stars
from Persson et al. (1998).  The observations {\bf were} carried out
with the jitter imaging technique and data analysis  {\bf was} performed
using the jitter routine from the ECLIPSE package.  The data
processing routines are described in Dotto et al. (2003) and Romon et
al. (2001). The obtained $V$-$J$, $V$-$H$ and $V$-$K$ colors are
reported in Table~\ref{tab:cols}.

\begin{figure}[h]
\centerline{\includegraphics[width=10cm]{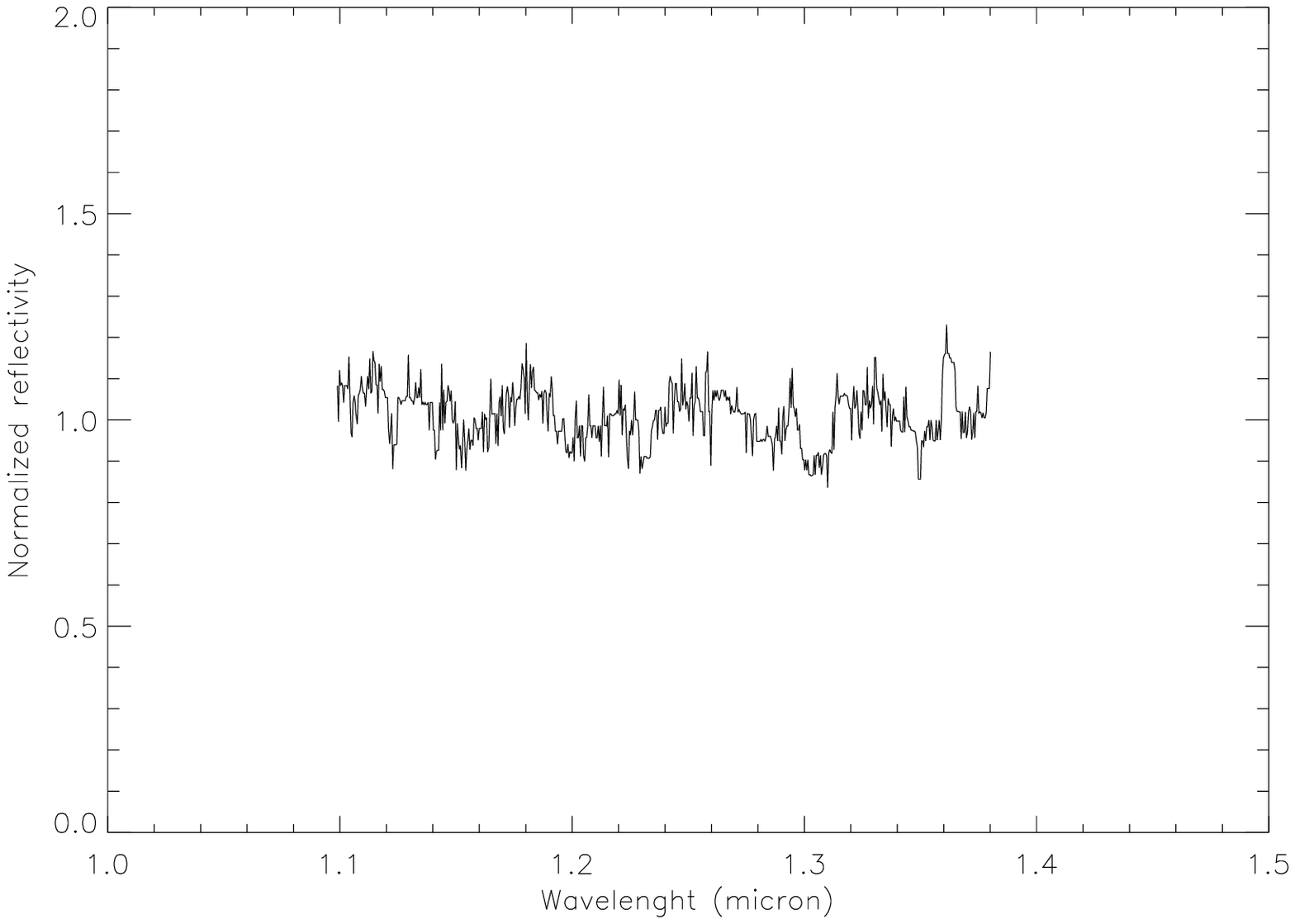}}
\centerline{\includegraphics[width=10cm]{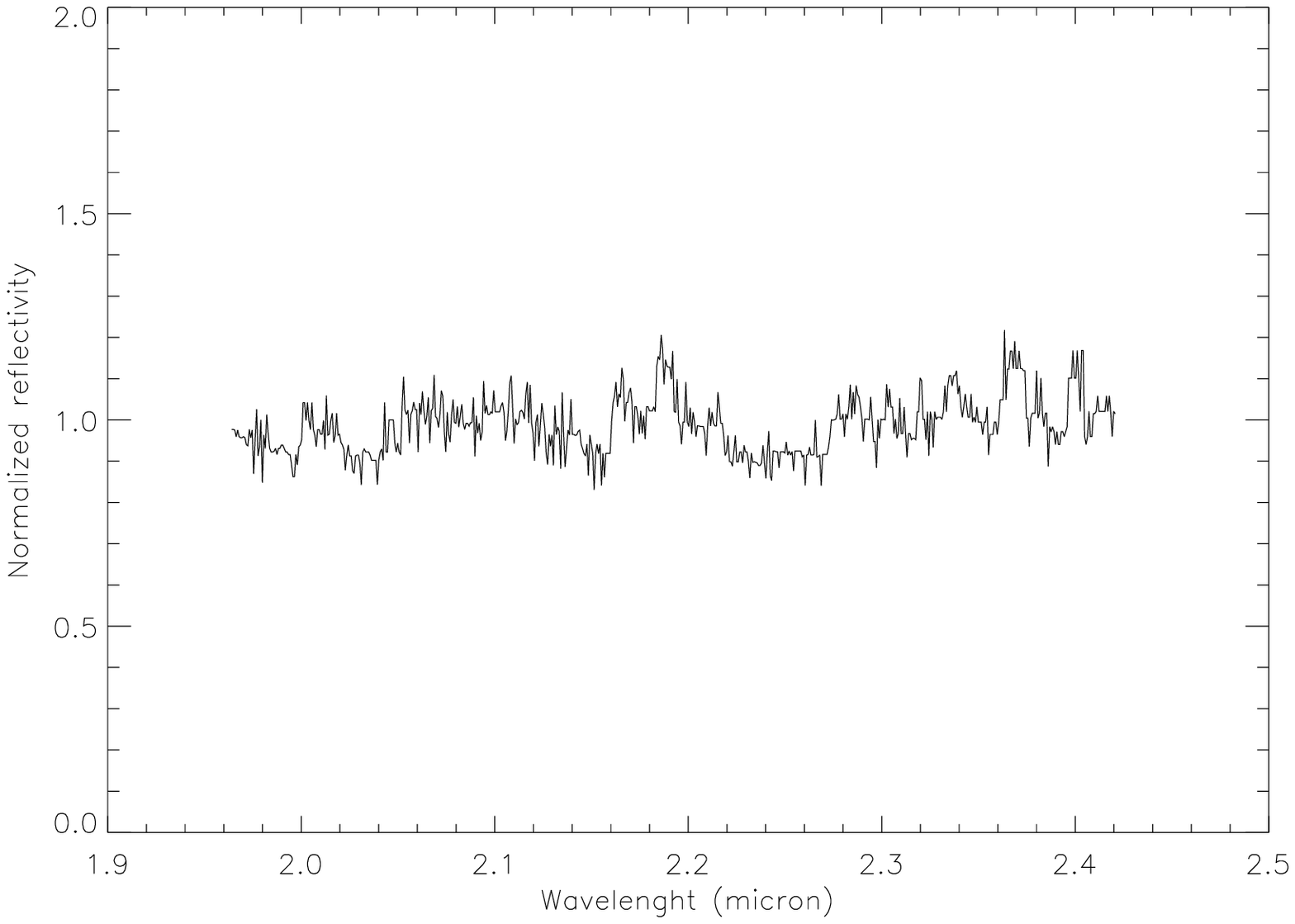}}
\caption{De Luise et al. -- Physical Investigation of PHA 2004VD17}
\label{fig:spc_J_K}
\end{figure}

Near-infrared spectroscopic observations were performed in the SW mode
(1 to 2.5 $\mu$m wavelength range). The ISAAC low resolution
spectroscopic mode, with a 2 arcsec wide slit and the grating at two
different central wavelengths corresponding to J and K bands, {\bf was used}
(Table~\ref{tab:texp_spc}).  The observations were done by nodding the
object along the slit by 10 arcsec between two positions A and B.  The
two averaged A and B images in each spectral range were subtracted
from each other. The A--B and B--A images were flat-fielded, corrected
for spatial and spectral distortion and finally combined with a
10-arcsec offset. The spectra were extracted from the resulting
combined images, and wavelength calibration was performed using a
xenon--argon lamp. The telluric absorption correction and the removal
of the solar contribution were {\bf obtained by dividing} the spectra of the
asteroid by {\bf the spectrum} of a solar analog star, observed just before and
after the object and at similar airmass.  The resulting spectra
are shown in Fig.~\ref{fig:spc_J_K}.


\section{Data analysis}

\indent
\textit{1. Polarimetry}

 [FIGURE 4] \\
 It is well known that asteroids exhibit a
well defined trend of the linear polarization versus the phase angle
(Fornasier et al. 2006b, and references therein). Two different
empirical relations {\bf make it possible} to determine the asteroid albedo thanks to
the knowledge of the minimum of the polarization curve ($P_{min}$)
or/and of the slope of the polarization curve near the inversion
angle. Primitive low albedo asteroids (e.g. C--types) present a
higher value of $P_{min}$ and a smaller inversion angle, as compared
for example to medium-high albedo objects (e.g. S-- or E--types).
Polarimetric albedos up to now available for the asteroid population
are in agreement with data derived from other techniques, including
direct measurements such as occultations and space missions.

\begin{figure}[h]
\centerline{\includegraphics[width=15cm]{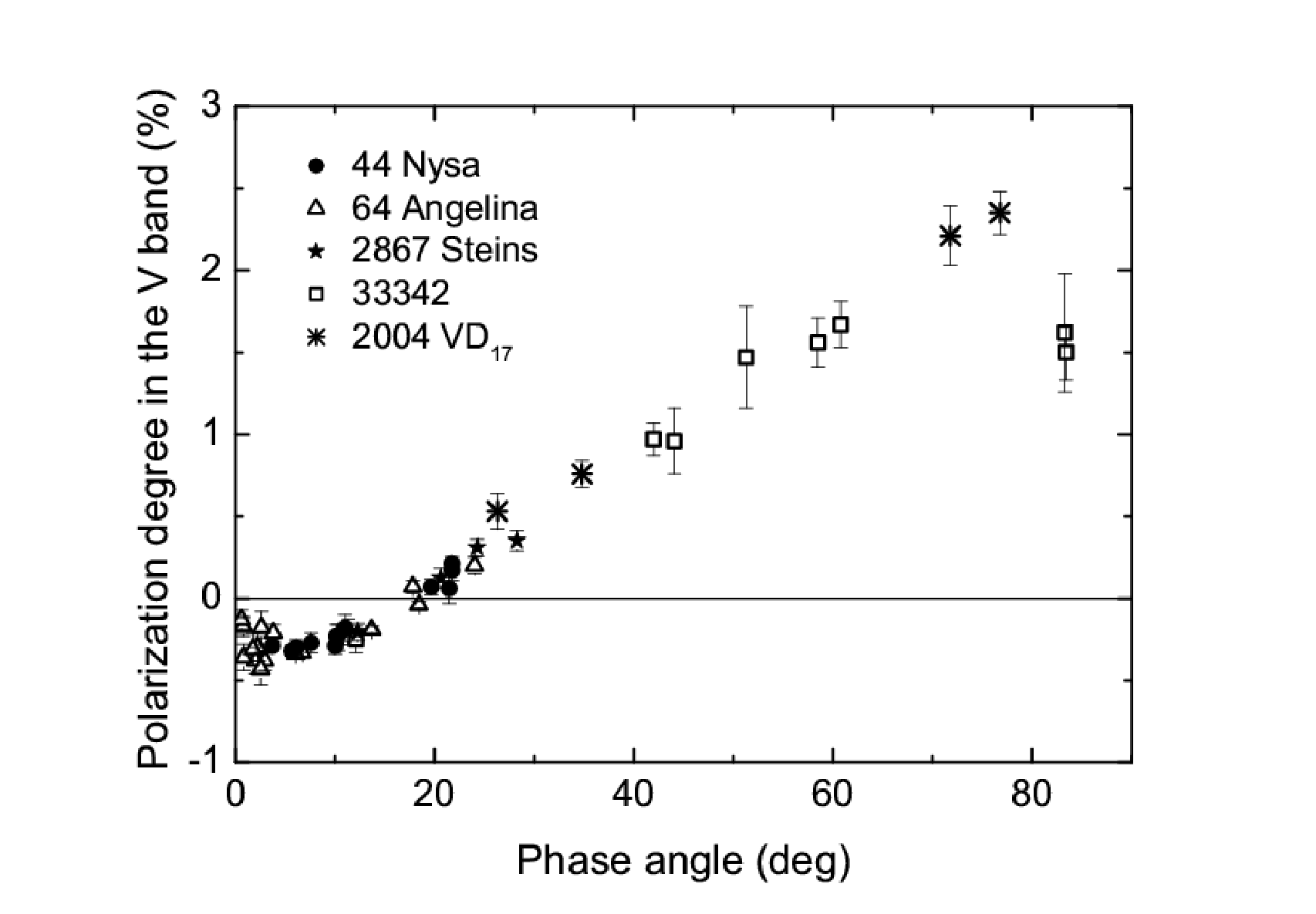}}
\caption{De Luise et al. -- Physical Investigation of PHA 2004VD17}
\label{fig:e_pol}
\end{figure}

To estimate the slope of the polarization phase curve at the inversion
angle in the V band, a linear fit to our data, weighted
according to their errors, {\bf was used} (see Fig.~\ref{fig:e_pol}). The use of a
linear fit is reasonable because of the small curvature of the
ascending branch of the asteroid polarization phase dependence (e.g.,
Zellner and Gradie 1976). 
{\bf The obtained slope is of 0.037$\pm$0.001 (\%/deg) in a good agreement with the mean slope of 0.041$\pm$ 0.007 (\%/deg) for the other four E-type asteroids observed so far (Zellner and Gradie 1976, Kiselev et al. 2002, Fornasier et al. 2006a)}.
{\bf The} relatively small
value of positive polarization degree of 2.35\% measured at phase
angle of 76.8$^{o}$ {\bf is another indication that the asteroid belongs
to the E class}. (144898) 2004 VD17 is the
second E-type NEO, together with (33342) 1998 WT24 (Kiselev et
al. 2002), for which polarimetric data are available. Comparison of
their polarimetric measurements is shown in Fig.~\ref{fig:e_pol},
which also includes  {\bf the} available observations of some main belt E-type
asteroids (Zellner and Gradie 1976; Rosenbush et al. 2005; Fornasier
et al. 2006a,b). Both NEOs and main belt asteroids show similar
polarization-phase angle dependences, suggesting similarity of their
microscopic surface properties. 
Our observations of (144898) 2004 VD17
are {\bf well} supplement to the available data, covering gaps in the
composite phase dependence of E-type asteroids shown in
Fig.~\ref{fig:e_pol}, and {\bf constraining a lower limit of the polarization maximum}. 
{\bf The data obtained for (144898) 2004 VD17 provides a maximum value of polarization $P_{max}\ge$ 2.35\%, higher than that one of (33342) 1998 WT24 (Kiselev et al. 2002). 
The phase angle of the polarization maximum for (144898) 2004 VD17 is $\alpha_{max} \ge 80^{o}$, while for (33342) 1998 WT24 is $\alpha_{max} \ge 75^{o}$. 
This discrepancy is likely connected with incertainties in polarimetric measurements rather than with differences in surface properties of the observed NEOs.}
To estimate
the albedo of 2004 VD17, the empirical correlation of
polarimetric slope vs. albedo, as described in Fornasier et al.
(2006a) for asteroid 2867 Steins {\bf was used}. These asteroids have exactly the
same values of the polarimetric slope assuming the same albedo.  The
corresponding albedo is 0.45$\pm$0.10, where the error {\bf accounts} both {\bf for}
the uncertainties on the slope and on the constants in the
slope-albedo relation.\\

\textit{2. Photometry and Spectroscopy}

\begin{figure}[h]
\centerline{\includegraphics{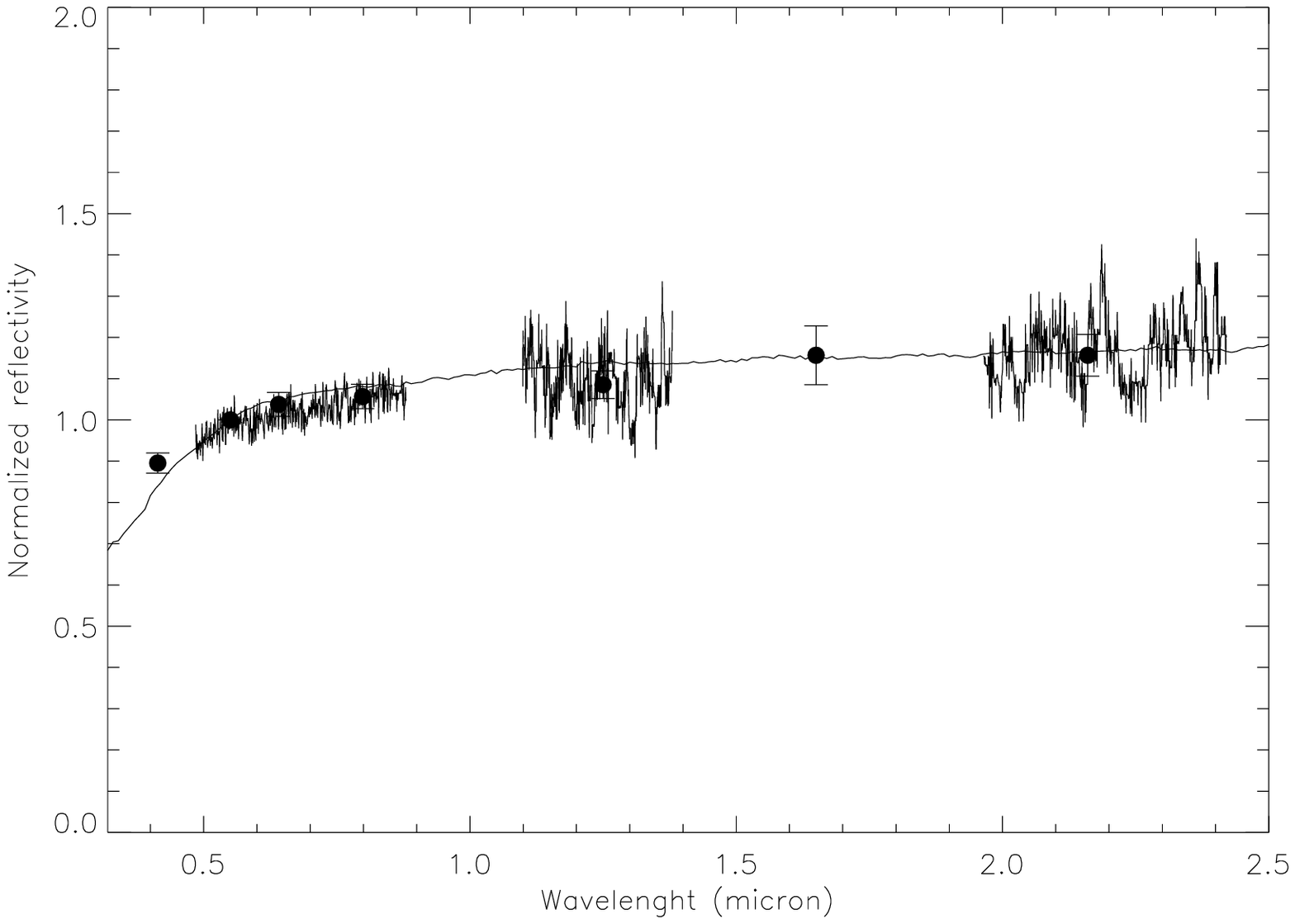}}
\caption{De Luise et al. -- Physical Investigation of PHA 2004VD17}
\label{fig:v_nir}
\end{figure}

[FIGURE 5] \\
The visible and
near-infrared colors reported in Table~\ref{tab:cols} were transformed
in reflectance and used to combine V, J, and K spectra.
Fig.~\ref{fig:v_nir} shows the entire spectrum of (144898) 2004 VD17,
from about 0.50 $\mu$m to about 2.4~$\mu$m{\bf. It} appears featureless
and almost flat, and compatible with {\bf the typical E-type
asteroids spectrum}.

In order to investigate the surface composition of this NEO, its
spectrum {\bf was} compared with a large sample of meteorite spectra
taken by the RELAB Public Spectroscopy Database
(http://www.planetary.brown.edu/relab/). The comparison {\bf was}
carried out by means of an automatic $\chi ^2$--test.  As shown in
Fig.~\ref{fig:v_nir}, a good match {\bf was} found with the
\textit{Mayo Belwa} aubrite meteorite.  This result gives a further
confirmation to our taxonomic classification of (144898) 2004 VD17,
since aubrites are widely believed to be the meteorite analogs of the
E-type asteroids. As the aubrite meteorites are mainly composed by
enstatite, the found match confirms the enstatitic nature of our
target. \\

\textit{3. Lightcurves}

\begin{figure}[h]
\centerline{\includegraphics[width=7cm,angle=90]{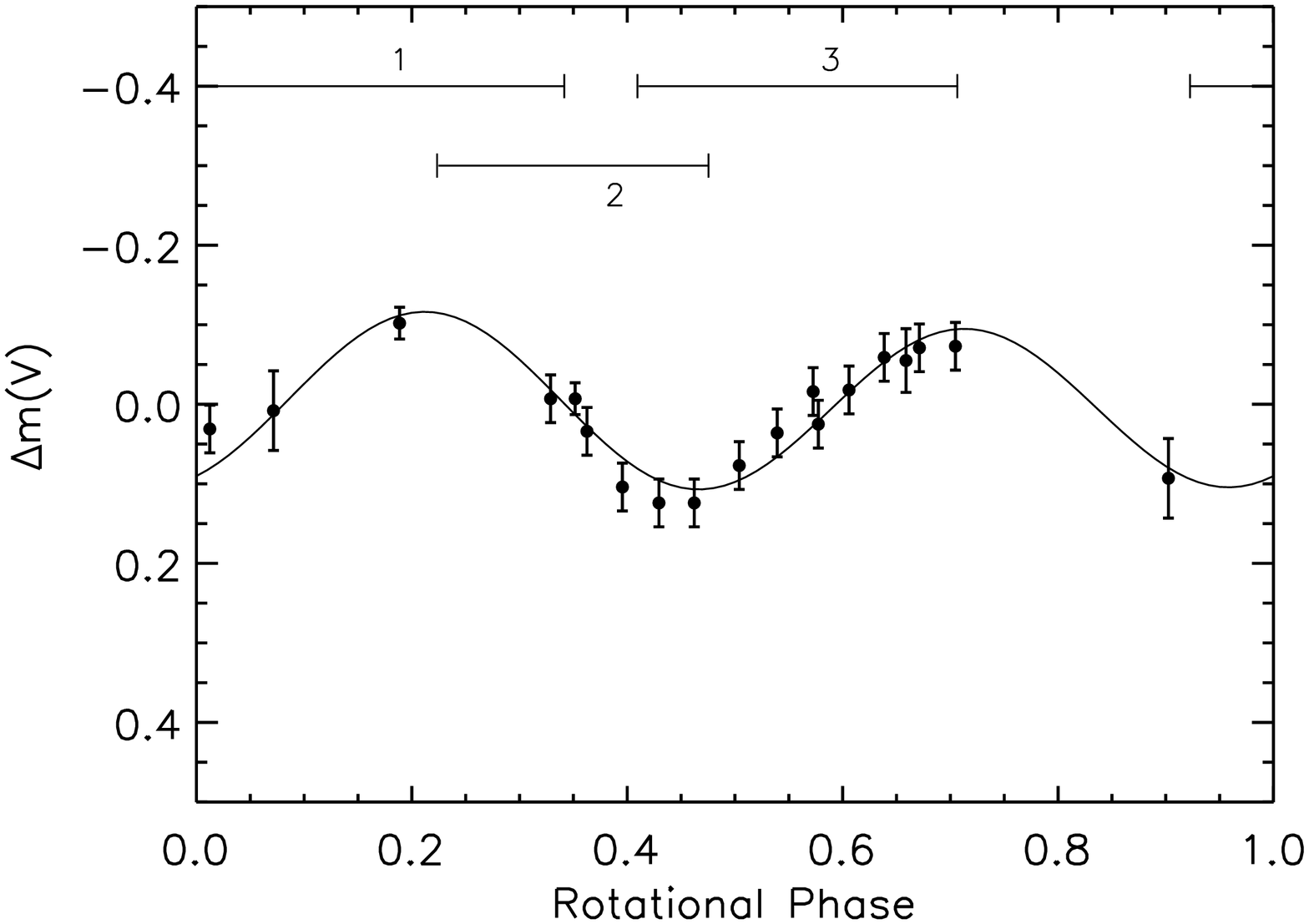}}
\centerline{\includegraphics[width=7cm,angle=90]{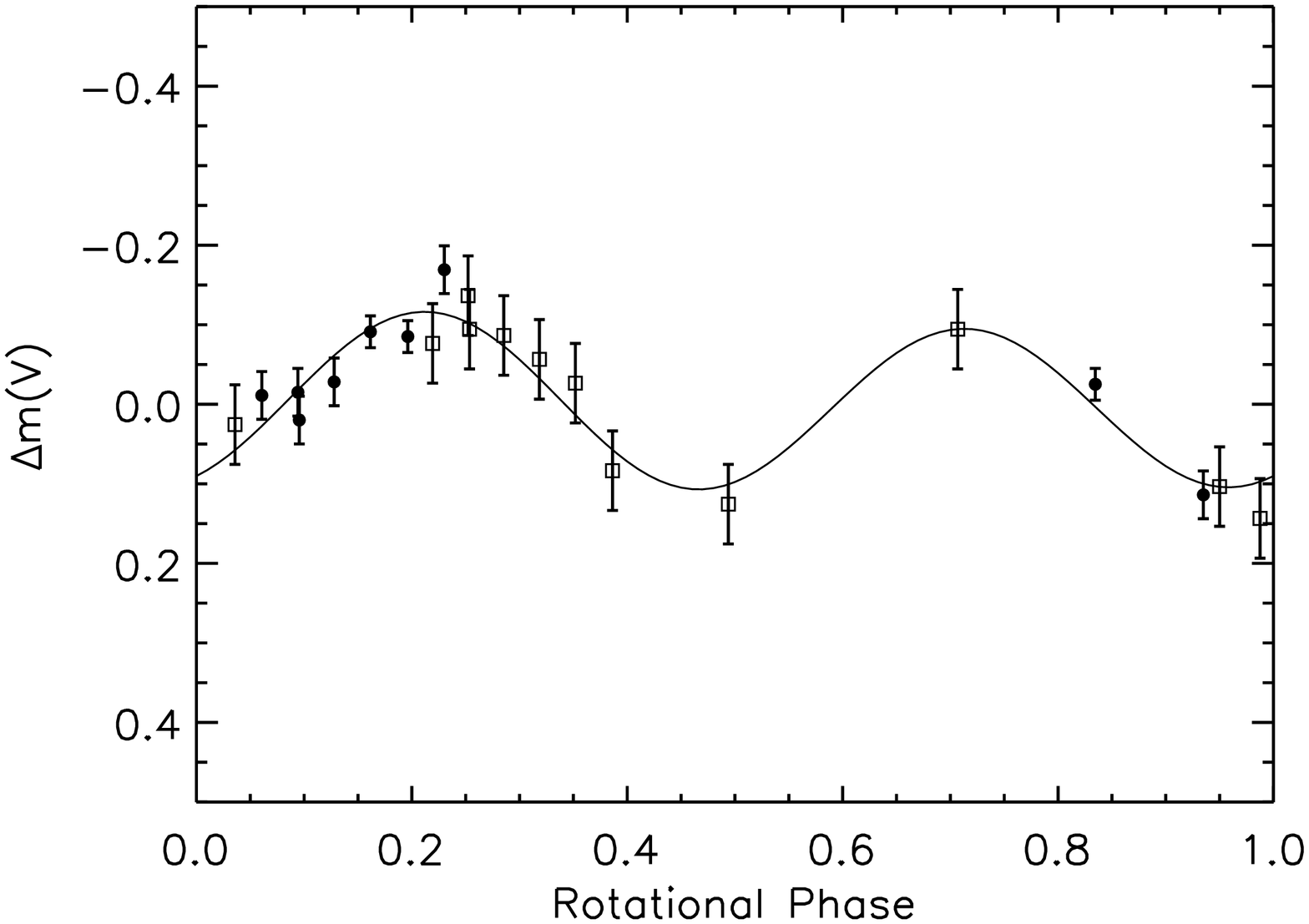}}
\caption{De Luise et al. -- Physical Investigation of PHA 2004VD17}
\label{fig:curV}
\end{figure}

\begin{figure}[h]
\centerline{\includegraphics[width=7cm,angle=90]{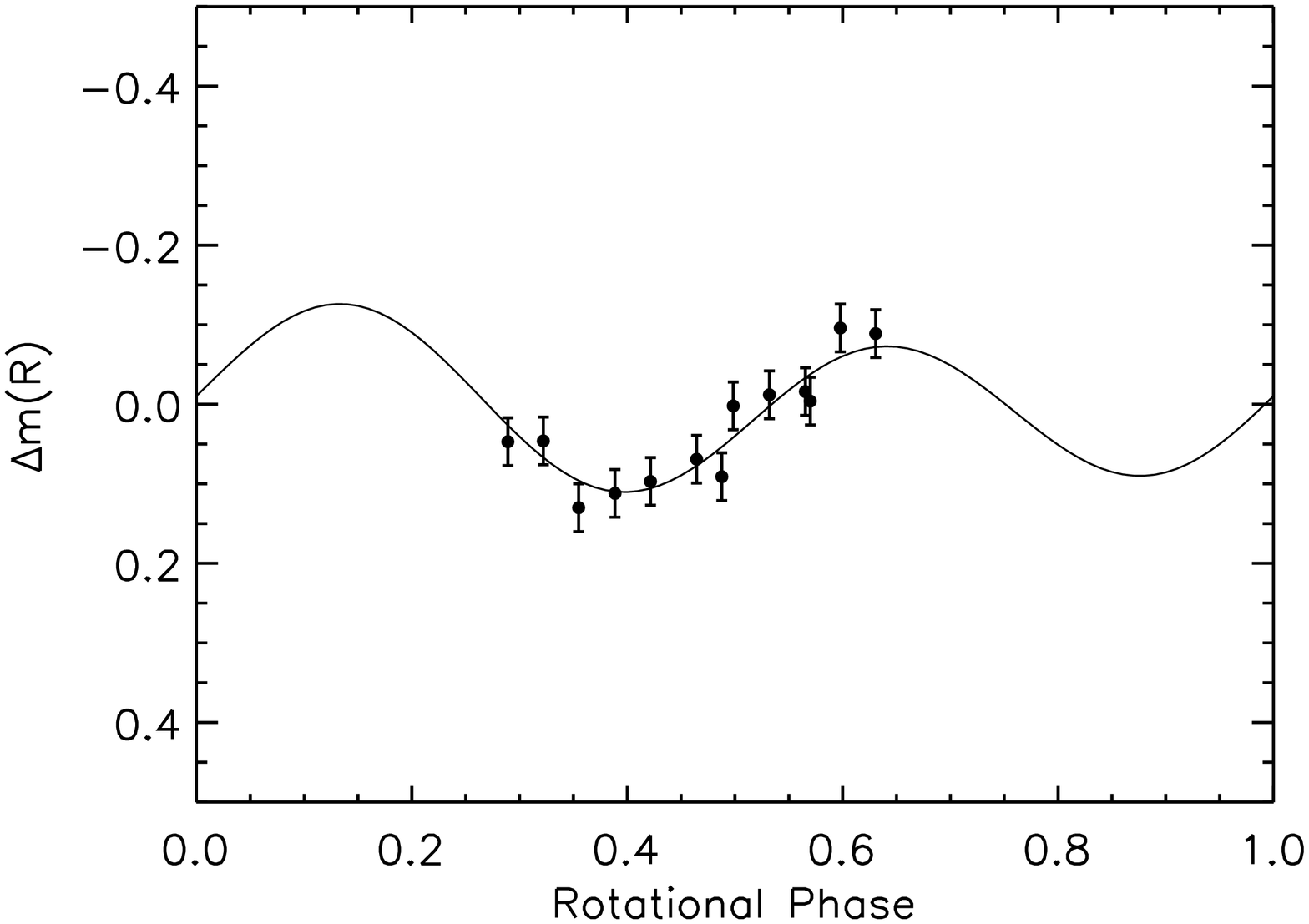}}
\centerline{\includegraphics[width=7cm,angle=90]{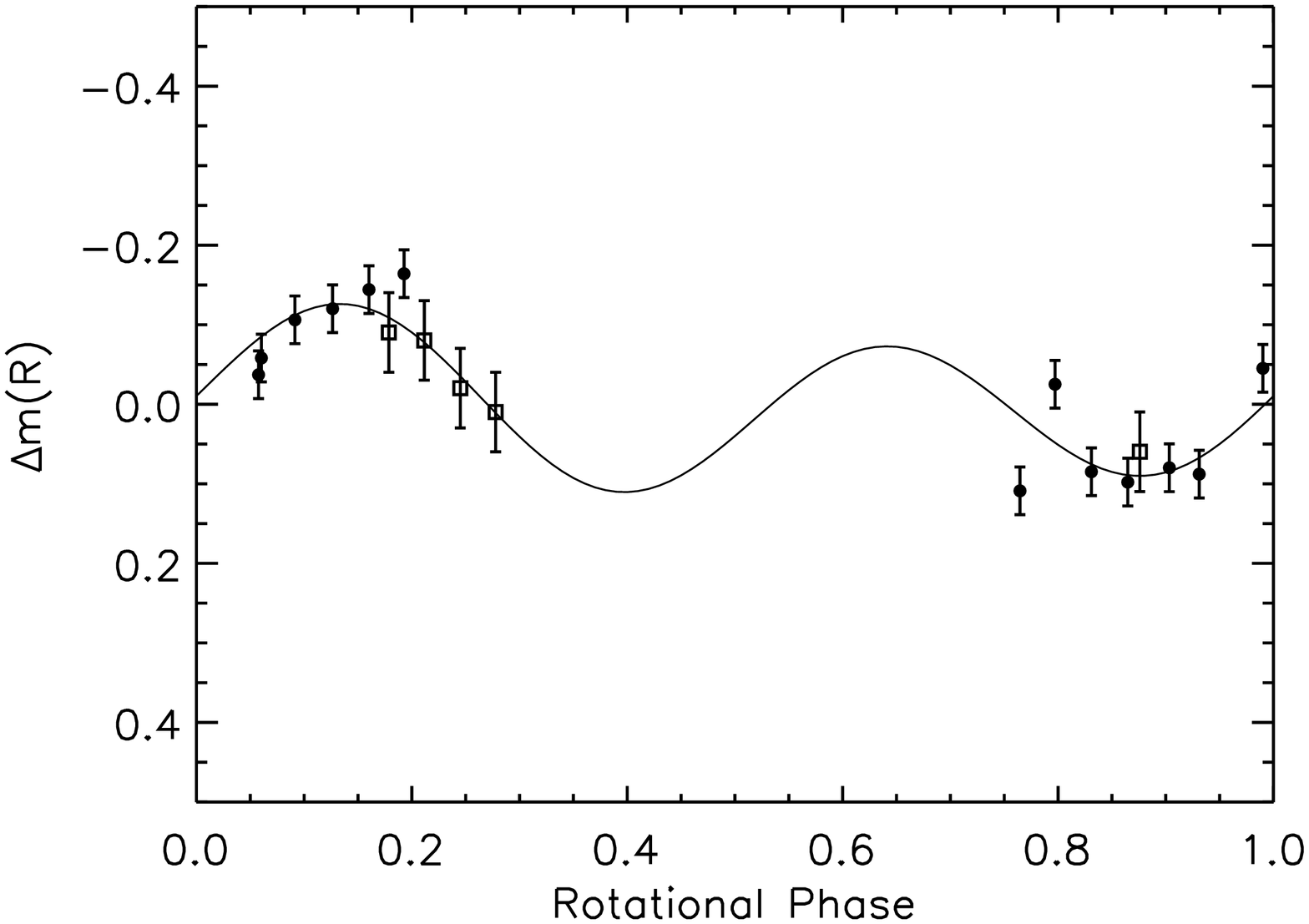}}
\caption{De Luise et al. -- Physical Investigation of PHA 2004VD17}
\label{fig:curR}
\end{figure}

[FIGURE 6 AND 7]\\
To determine the synodic rotational period
of (144898) 2004 VD17, a Fourier analysis of our
photometric data set, as described in Harris et al. (1989)
{\bf was performed}. {\bf The rotational period of 2004 VD17 resulted} $P_{syn} = 1.99 \pm 0.02$
hours. The obtained V and R lightcurves are shown in Figures
\ref{fig:curV} and \ref{fig:curR} respectively. Their behaviors {\bf make it possible} to exclude a binary nature.

The lightcurve amplitude is $0.21 \pm 0.02$~mag, {\bf giving} a
lower limit of the semimajor axis ratio {\bf of} $a/b \geq 1.21 \pm 0.02$.
On the basis of these results, 2004 VD17 is a fast rotator asteroid
(FRA) with an ellipsoidal shape: this makes it a peculiar object
since it is widely believed that fast spin rates are coupled with
quite spheroidal structures (Pravec and Harris 2000). It could be
speculated that the repeated recent close encounters with the terrestrial 
planets (Scheeres et al. 2004) and/or the non gravitational perturbations 
(e.g. Yorp effect, {\bf Lowry et al. 2007}), related to its non spherical shape and small semimajor axis,
brought the spin rate of our target to the present high value.

In the Fig.~\ref{fig:curV} the coverage of the three visible spectra
shown in Fig.~\ref{fig:3vis} is reported. Since we found the same
spectral behavior for about the 75\% of the whole rotational phase,
we can {\bf argue} that (144898) 2004 VD17 has a quite homogeneous surface
composition.

\section{Discussion}

We applied the Bowell et al. (1989) procedure to our V photometric data together with all the astrometric V magnitudes, archived from 1st March 2006 to 1st June 2006 in the \textit{NEODyS}\footnote{{\tt
http://newton.dm.unipi.it/cgi-bin/neodys/neoibo}} database
(http://newton.dm.unipi.it/cgi-bin/neodys/neoibo), to estimate the absolute magnitude of (144898) 2004 VD17, obtaining $H=18.9$. 
Hence, we estimated the diameter of our target using the empirical 
relationship
\begin{equation}
 D = \frac{ 1329 \cdot 10^{-H/5}}{\sqrt{p_v} }
\end{equation}
with a visual albedo of $p_v=0.45$, obtaining a diameter of $D\sim320$ m.

Using the {\bf measured} synodic period ($P_{syn}$) and the obtained
lightcurve amplitude ($\Delta m$), {\bf from} the Eq. (7) of Pravec and Harris
(2000):
\begin{equation}
\rho \simeq \left(\frac{3.3}{P_{syn}}\right)^2 \cdot (1+\Delta m)
\end{equation}
\noindent
we can estimate the density of (144898) 2004 VD17 {\bf assuming}
a body with no tensile strength (rubble pile).  The resulting value is
$\rho \sim 3.36 $~g/cm$^3$, that would be an extremely large value
with respect {\bf to} the currently known asteroids' densities. Note that our
target, with its non negligible ellipsoidal shape, would lie slightly
beyond the border between ``rubble-piles'' and ``monolithic'' bodies,
as shown in Fig. 8 of Pravec and Harris (2000). Following their
analysis, with $D\sim 320$ m, $a/b \sim 1.2$ and $P_{syn} \sim 2$
hours, (144898) 2004 VD17 would be a peculiar, quite large fast
rotator, with a larger-than-average non-spherical shape, calling for a
very high density value to withstand the rotational accelerations.
Moreover, with about 12 revolutions per day, in the classical
plot ``Diameter vs. Spin rate'', our target would lie above the
so-called ``rubble-pile spin barrier'', but to the right of the $0.2$
km line, usually discriminating between ``small monoliths'' and
``rubble-piles''.  This seems to suggest a more robust structure.
(144898) 2004 VD17 could be a monolith, maybe partially fractured
(Richardson et al. 2002), even if the lack of any information about
its porosity does not make possible to constrain its nature.  On the other
hand, recent works by Holsapple (2002, 2003) pointed out the possible
role of cohesive forces in determining the strength of an asteroid. In
particular, for asteroids of the size of our target where the
gravitational pressures are small, the cohesion forces could become
important so that we could speak about a ``strength regime'' (as it is
customary in impact cratering studies). Note that an extremely low
cohesion (compared to terrestrial rocks), of the order of a few parts
in $10^4$, is enough to hold a non-spherical fast rotating body, such
as (144898) 2004 VD17, together.  An asteroid model with a low
cohesion gives a bound on spin rate above the classical barrier,
leading to a smooth transition between the data from small fast
rotators to the larger bodies (where the gravity pressure
dominates). E.g., in Fig. 3 of Holsapple (2003), (144898) 2004VD17
would fall within the new spin limits set by the $5 \times 10^4$
dyne/cm$^2$ curve.  In other words, it is not necessary to refer to
the category of ``monoliths'' to account for the fast rotation rate of
our target, and a rubble pile structure with a neglegible strength
would suffice.

Considering diameter and density of our NEO, we can obtain an
estimation of the impact energy, $E_{imp}$ (Chapman and Morrison
1994; Chesley et al. 2002).  The impact velocity for the 2102
encounter is evaluated as $V_{imp} =21.36$ km/s (see, e.g., NEODyS).
Assuming a range of densities from $\rho = 2$ to $\rho = 3.36$~g/cm$^3$, $E_{imp}$ varies from $\approx 7.8 \times 10^{18}$ J to
$\approx 1.3 \times 10^{19}$~J, corresponding to $1871$ and $3143$ MT,
respectively.  Note that these values, while $4.5$ to $7.5$ times lower than
those used by NEODyS to assess the risk, are still about two orders of
magnitudes larger than the Tunguska event.

We rerun, with the methods of Milani et al. (2005), the impact
risk estimation for the 2102 possible impact of (144898) 2004 VD17,
which still has a small but non negligible probability, estimated at
$4.2\times 10^{-7}$ on the NEODyS risk page.  We
have used as mass value the upper limit of the range estimated above,
rather than the  {\bf mass} computed on the basis of an ``average'' NEO
albedo, as it is done when no data are available (Chesley et al.
2002). 

If the risk is measured as \emph{expected impact energy} (impact
energy times probability), the value changes from $5.9$ KT to
$1.3$ KT, in proportion to the mass. The other useful value is given by the Palermo Scale. The PS value has changed with the
revised mass by $-0.52$ (corresponding to a factor $3.3$ in the
probability ratio); this factor is somewhat smaller than the one for
the mass because the background probability does not scale in inverse
linear relationship with the impact energy.  The PS change {\bf was}
from $-3.52$ to $-4.04$, thus this case is now beyond the
``psycological barrier'' set at PS $=-4$ by the warnings posted on the
NEODyS risk page.  Nevertheless, because the orbit is now very well
determined and no radar observations are possible for a long time,
(144898) 2004 VD17 could remain on the ``risk pages'' of NEODyS and of
the corresponding JPL Sentry system, although with a low risk level,
for decades. 

The results {\bf we obtained} for 2004 VD17 suggest some comments on our estimation of the
impact risk and on the discovery completeness of the NEO population.
Recent observations made possible to measure the albedo of the two NEOs with most stable Virtual Impactors,
(99942) Apophis (Cellino et al. 2007) and, in this work, (144898) 2004 VD17. The measured albedos are {\bf much} larger than the mean value of 0.1
usually adopted to convert the observed magnitudes in sizes.
This resulted in an initial size overestimation, with a corresponding
overestimation of the expected impact energy and Palermo Scale{\bf.  This
overestimation} could occur systematically for Virtual Impactors, that are {\bf usually} quite
small (the impact probability of large objects is by far smaller).
{\bf It is evident that, at a given magnitude cut-off, discovery surveys 
detect a larger fraction of higher
albedo bodies, with respect to the fraction of bodies in the population corresponding, for the average
albedo, to a given size. Therefore, the NEOs having virtual impactors could be biased in favour of high
albedo objects.}


\section{Conclusions}

During April and May 2006, a visible and near--infrared
photometric and spectroscopic survey, as well as polarimetric
observations of the NEO (144898) 2004 VD17, {\bf were carried out} at the ESO-Chile
telescopes.
              
With our survey, a complete analysis of (144898) 2004 VD17 {\bf was obtained}:
\begin{itemize}
\item[\it i)] 
on the basis of our polarimetric data an
albedo of $p_v\sim0.45 \pm 0.10$  {\bf was computed} and a diameter of about
$D\sim320$ m {\bf was estimated};
\item[\it ii)] 
{\bf from the} visible photometry the rotational
period of $1.99 \pm 0.02$ hours {\bf was computed}. The lightcurve
analysis gave a {\bf lower} limit of the semimajor axis {\bf ratio} 
$a/b \geq 1.21 \pm
0.02$. 2004 VD17 is therefore a fast rotator object with an unexpected
ellipsoidal shape;
\item[\it iii)] 
photometric images and lightcurve analysis
{\bf found no} companions {\bf nor} cometary activity;
\item[\it iv)] 
visible spectra obtained at different rotational phases
are very similar each other, suggesting an homogeneous surface
composition for this object;
\item[\it v)] 
{\bf from} the flat and featureless visible and near--infrared
spectra, (144898) 2004 VD17 {\bf was classified} as an E-type asteroid and
analogies with the spectra of the aubrite meteorites {\bf were found}.  Our
taxonomic classification has been confirmed also by the high value of
the polarimetric albedo.
\end{itemize} 
On the basis of our results, (144898) 2004 VD17 is therefore a very
peculiar object.  According to its density, diameter and rotational
period, it is close to the limit estimated for the disruption of the
``rubble-pile" structure, making hard to distinguish it as an
aggregate, a monolith or a fractured monolith. The rubble pile
structure is still possible if a strength regime is considered with a
cohesion of a few times $10^4$ dyne/cm$^2$ (Holsapple 2002, 2003).  
Further investigations on the porosity and the relative tensile strength 
are needed to better constrain the internal structure of this object.

(144898) 2004 VD17 has still a possibility of impacting the Earth (in
the year 2102){\bf,} compatible with the available astrometry.
{\bf This} risk cannot be easily removed by additional astrometry. Thus it
{\bf was} important to reduce the risk estimate by a factor $4.5$
in the expected impact energy and by a factor $3.3$ in the ratio of
the probability of this event to the probability of an event of the
same energy from the background population. Besides its scientific
interest, (144898) 2004 VD17 is a dangerous object. {\bf Observing it
had} the additional value of contributing {\bf to decrease} the
estimated risk.\\


{\bf Acknowledgments}

The present work is part of the project funded by PRIN-MIUR 2004 ``{\it The Near Earth Objects as an opportunity to understand physical and dynamical
properties of all the solar system small bodies}" (PI: A. Milani).  In
particular, this investigation is also the result of a discussion held
at the 2005 plenary meeting of this project. \\
FDL wants to thank Massimo Dall'Ora, Simone Antoniucci and Luca
Calzoletti for the helpful discussions about the NIR data analysis,
and Germano D'Abramo for his kind and quick help.

\bigskip


{\bf References}

Binzel, R.P., 2000. The Torino Impact Hazard Scale. PSS 48, 297-303

Bowell, E., Hapke, B., Dominigue, D., Lumme, K., Peltonimei, J., Harris, A. W., 1989. Application of photometric models to asteroids. In: R. P. Binzel, T. Gehrels and S. M. Matthews (Eds.), Asteroids II  524-556

In: W. F. Bottke Jr., A. Cellino, P. Paolicchi, and R. P. Binzel (Eds.), Asteroids III. Univ. Arizona Press, Tucson, pp. 501-515

Cellino A., Delb$\grave{o}$ M. and Tedesco E. F., 2007. Albedo and size of (99942) Apophis from polarimetric observations. In:  Milani A., Valsecchi G.B. and Vokrouhlick´y D. (Eds.), Near Earth Objects, our Celestial Neighbors: Opportunity and Risk. Proceedings IAU Symposium No. 236. Cambridge University Press, pp 451-454.

Chapman, C.R., Morrison, D.C., 1994. Impact on the Earth by Asteroids and Comets: Assessing the Hazard. Nature 367, 33-40

Chesley, S.R., Chodas, P.W., Milani, A., Valsecchi, G.B., Yeomans, D.K., 2002. Quantifying the Risk Posed by Potential Earth Impacts. Icarus 159, 423-432 

Dotto, E., Barucci, M. A., Leyrat, C., Romon, J., de
Bergh, C., Licandro, J., 2003. Unveiling the nature of 10199 Chariklo: near-infrared observations and modeling. Icarus 164, 122-126.

Dotto, E., Fornasier, S., Barucci, M. A., J. Licandro, Boehnhardt, H., Hainaut, O., Marzari, F., de Bergh, C., De Luise, F., 2006. The surface composition of Jupiter Trojans: Visible and near-infrared survey of dynamical families. Icarus 183, 420--434,

Fornasier, S., Belskaya, I., Fulchignoni, M., Barucci, M. A., Barbieri, C., 2006a. First albedo determination of 2867 Steins, target of the Rosetta mission. A\&A 449, L9-L12 

Fornasier, S., Belskaya, I. N., Shkuratov, Yu. G., Pernechele, C., Barbieri, C., Giro, E., Navasardyan, H., 2006b. Polarimetric survey of asteroids with the Asiago telescope. A\&A 455, 371-377

Harris, A.W., Young, J. W., Bowell, E., Martin, L. J., Millis, R. L., Poutanen, M., Scaltriti, F., Zappala, V., Schober, H. J., Debehogne, H., Zeigler, K. W., 1989. Photoelectric observations of asteroids 3, 24, 60, 261, and 863. Icarus 77, 171-186

Holsapple, K. A., 2002. Speed limits of rubble pile
asteroids: Even fast rotators can be rubble piles. September, 2002,
Workshop on Scientific Requirements for Mitigation of
Hazardous Comets and Asteroids, Washington

Holsapple, K. A., 2003. Could fast rotator asteroids be rubble piles?
34th Annual Lunar and Planetary Science Conference, 
March, 2003. League City, Texas, abstract no. 1792.

Kiselev N.N., Rosenbush V.K., Jockers, K., Velichko, F. P., Shakhovskoj, N. M., Efimov, Yu. S., Lupishko, D. F., Rumyantsev, V. V., 2002. Polarimetry of near-Earth asteroid 33342 (1998 WT24). Synthetic phase angle dependence of polarization for the E-type asteroids. Proceeding ACM 2002, Berlin, ESA - SP-500, 887-890.

Landolt, A.U., 1992. UBVRI photometric standard stars in the magnitude range 11.5-16.0 around the celestial equator. AJ 104, 340-371, 436-491

{\bf Lowry, S.C., Fitzsimmons, A., Pravec, P., Vokrouhlick$\grave{y}$, D., Boehnhardt, H., Taylor, P.A., Margot, J.L., Gal$\grave{a}$d, A., Irwin, M., Irwin, J., Kusnir$\grave{a}$k, P., 2007. Science 316, 272L}

Milani, A., Chesley, S.~R., Sansaturio, M.~E., Tommei, G., Valsecchi,
  G. B., 2005. Nonlinear impact monitoring: Line Of Variation searches
  for impactors. Icarus 173, 362--384

Persson, S. E., Murphy, D. C., Krzeminski, W., Roth, M., Rieke, M. J., 1998. A New System of Faint Near-Infrared Standard Stars. AJ 116, 2475-2488 

Pravec, P. and Harris, A. W., 2000. Fast and Slow Rotation of Asteroids. Icarus 148, 12-20 

Richardson, D. C., Leinhardt, Z.M., Melosh, H.J., Bottke, W.F. Jr., Asphaug, E., 2002. Gravitational Aggregates: Evidence and Evolution. In: W. F. Bottke Jr., A. Cellino, P. Paolicchi, and R. P. Binzel (Eds.), Asteroids III. Univ. Arizona Press, Tucson, pp. 501-515

Romon, J., de Bergh, C., Barucci, M. A., Doressoundiram, A., Cuby, J.-G., Le Bras, A., Dout$\grave{e}$ S., Schmitt, B., 2001. Photometric and spectroscopic observations of Sycorax, satellite of Uranus. A\&A 376, 310--315

Rosenbush, V. K., Kiselev, N. N., Shevchenko, V. G., Vasilij G., Jockers, K., Shakhovskoy, N. M., Efimov, Y. S., 2005. Polarization and brightness opposition effects for the E-type Asteroid 64 Angelina. Icarus 178, 222--234.

Scheeres, D. J., Marzari, F. and Rossi, A., 2004. Evolution of NEO rotation rates due to close encounters with Earth and Venus. Icarus 170, 312--323. 

Zellner, B. and Gradie J., 1976. Minor planets and related objects. XX - Polarimetric evidence for the albedos and compositions of 94 asteroids. ApJ 81, 262--280.

\newpage


{\bf Tables}

\begin{table*}[h]
{\small
       \begin{center}
       \caption{Orbital parameters of (144898) 2004 VD17}
        \label{tab:orbpar}
\begin{tabular}{ l c } \hline
 Perihelion Distance (AU) & 0.6203  \\
 Aphelion Distance (AU)   & 2.3960   \\
 Eccentricity             & 0.5887   \\
 Inclination (degree)     & 4.223    \\
 Ascending Node (degree)  & 224.239  \\
 Argument of perihelion (degree)  & 90.686  \\
\hline

\end{tabular}
\end{center}}
\end{table*}

\begin{table*}[h]
{\small
       \begin{center}
       \caption{Observational circumstances: heliocentric and geocentric distances ($r$ and $\Delta$) and phase angle ($\alpha$) are referred at 0UT of each night.}
        \label{tab:surv}
\begin{tabular}{ |l|c|c|c|c|c|c|c|} \hline
\hline
Date  &  Telescope & Instrument & Range & Technique(s) &   \textit{r}   & \textit{$\Delta$} & \textit{$\alpha$} \\
&&&&& (AU) & (AU)   & (degree)  \\
\hline
16 Apr 06 & VLT-UT2 & FORS1 & Vis & Polarimetry & 1.426 & 0.497 & 26 \\
23 Apr 06 & VLT-UT2 & FORS1 & Vis & Polarimetry & 1.365 & 0.485 & 35 \\
01 May 06 & NTT     & EMMI  & Vis & IMG + SPC   & 1.292 & 0.476 & 44 \\
21 May 06 & NTT     & EMMI  & Vis & IMG + SPC   & 1.098 & 0.460 & 67 \\
22 May 06 & NTT     & EMMI  & Vis & IMG + SPC   & 1.088 & 0.459 & 68 \\
24 May 06 & VLT-UT1 & ISAAC & NIR & IMG + SPC   & 1.068 & 0.456 & 71 \\
25 May 06 & VLT-UT1 & ISAAC & NIR & IMG + SPC   & 1.057 & 0.454 & 72 \\
25 May 06 & VLT-UT2 & FORS1 & Vis & Polarimetry & 1.057 & 0.454 & 72 \\
28 May 06 & VLT-UT2 & FORS1 & Vis & Polarimetry & 1.018 & 0.448 & 77 \\

\hline

\end{tabular}
\end{center}}
\end{table*}

\begin{table*}[h]
   \caption{Results for the (144898) 2004 VD17 polarimetric observations in the V filter. The position angles of the scattering plane ($\phi$) {\bf were} taken from the JPL ephemeris service (http://ssd.jpl.nasa.gov/cgi-bin/eph).}
\begin{center}
\label{tab:polar}
\begin{tabular}{|l|c|c|c|c|c|c|c|} 
\hline
Date & UT$_{start}$ &   $\phi$ ($^{o}$) & P (\%) & $\theta$   & P$_r$ (\%) & $\theta_r$ ($^{o}$)  \\ \hline
16 Apr 06 & 02:09 & 105.1 & 0.53$\pm$0.11 & 14.1$\pm$5.9 & +0.53$\pm$0.11 & -1.1$\pm$5.9  \\
23 Apr 06  & 01:27 & 107.8 & 0.76$\pm$0.08 & 21.0$\pm$3.0 & +0.76$\pm$0.08 & 3.1$\pm$3.0 \\
25 May 06  & 00:05 & 109.9 & 2.21$\pm$0.18 & 19.7$\pm$2.3 & +2.21$\pm$0.18 & -0.2$\pm$2.3  \\
28 May 06  & 23:32 & 109.6 & 2.35$\pm$0.13 & 20.0$\pm$1.6 & +2.35$\pm$0.13 & 0.4$\pm$1.6  \\
\hline
      \end{tabular}
       \end{center}
       \end{table*}

\begin{table*}[h]
{\small
       \begin{center}
       \caption{(144898) 2004 VD17 colors}
        \label{tab:cols}
\begin{tabular}{|c|c|c|c|c|c|} \hline
\hline
 B-V & V-R & V-I & V-J & V-H & V-K \\
\hline
 

0.76$\pm$0.03 & 0.40$\pm$0.03 & 0.75$\pm$0.03 & 1.16$\pm$0.04 & 1.15$\pm$0.08 & 1.56$\pm$0.06 \\
 

\hline
\end{tabular}
\end{center}}

\end{table*}

\newpage

\begin{table*}[h]
{\small
       \begin{center}
       \caption{Observational circumstances for visible and near-infrared spectroscopy}
        \label{tab:texp_spc}
\begin{tabular}{ |l|c|c|c|c|c|c|c| } \hline
\hline

Date     & UT$_{start}$ &Grism & Slit     & T$_{exp}$ & n$_{exp}$ & airmass & Solar Analog   \\
          &              &       & (arcsec) & (s)     & 	     &         &  (airmass)  \\
\hline
01 May 06 & 01:14 & \#1   &  5.0     & 1800    & 3$\times$600s & 1.27 & SA102-1081 (1.20) \\
20 May 06 & 23:26 &  \#1   &  5.0     & 1200    & 2$\times$600s & 1.34 & SA102-1081 (1.16) \\
21 May 06 & 23:41 &  \#1   &  5.0     & 1200    & 2$\times$600s & 1.40 & SA102-1081 (1.17) \\

\hline
\hline

Date     & UT$_{start}$ & Band           & Slit     & T$_{exp}$ & n$_{exp}$ & airmass & Solar Analog    \\
          &  & ($\mu$m)       & (arcsec) & (s)     & 	      &         &    (airmass)\\
\hline
25 May 06 & 00:05 & 1.10$\div$1.30 &  2.0     & 7200    & 3$\times$2400s & 1.15 & SA102-1081 (1.23)     \\
25 May 06 & 01:51 & 1.84$\div$2.56 &  2.0     & 2520    & 1$\times$2520s & 1.14 & SA102-1081 (1.21)     \\

\hline

\end{tabular}
\end{center}}
\end{table*}


\newpage

{\bf Figure captions}

FIGURE 1: Single night lightcurves of (144898) 2004 VD17 corrected for light-time. For V filter: (a) 1st May, (b) 21st May, (c) 22nd May; for R filter: (d) 1st May, (e) 21st May, (f) 22nd May.

FIGURE 2: Visible spectra of (144898) 2004 VD17 normalized at 0.55 $\mu$m. The three spectra {\bf are} shifted by 0.5 for clarity

FIGURE 3: Near-infrared spectra of (144898) 2004 VD17: (\textit{up}) J filter, normalized at 1.25$\mu$m; (\textit{down}) Ks filter, normalized at 2.16 $\mu$m.

FIGURE 4: Polarization degree in the V band versus phase angle of (144898) 2004 VD17 and Near Earth and Main Belt E--type asteroids from Kiselev et al. (2002) and Fornasier et al. (2006a).

FIGURE 5: (144898) 2004 VD17 visible and near-infrared combined spectrum, normalized at 0.55 $\mu$m. The continuous line represents the spectrum of the Mayo Belwa aubrite meteorite.

FIGURE 6: (144898) 2004 VD17 composed lightcurves in V filter, folded with a period of 1.99 hours: \textit{(up)} 1st May night: $\Delta m=0$ corresponds to a mean magnitude of 19.32; the coverage of the spectra obtained on 1st May \textit{(1)}, 21st May \textit{(2)} and 22nd May \textit{(3)} is shown on the top; \textit{(down)} 21st May (black dots) and 22nd May (white squares): $\Delta m=0$ corresponds to a mean magnitude of 19.68. The lightcurve zero point is at 0UT of 12th May.

FIGURE 7: (144898) 2004 VD17 composed lightcurves in R filter, folded with a period of 1.99 hours: \textit{(up)} 1st May night: $\Delta m=0$ corresponds to a mean magnitude of 18.93; \textit{(down)} 21st May (black dots) and 22nd May (white squares): $\Delta m=0$ corresponds to a mean magnitude of 19.29. The lightcurve zero point is at 0UT of 13th May.


\newpage

{\bf Figures}

\end{document}